\documentclass[12pt]{article}
\begin{document}
\baselineskip 18pt
\newcommand{\Tr}{\mbox{Tr\,}}
\newcommand{\Dirac}{/\!\!\!\!D}
\newcommand{\beq}{\begin{equation}}
\newcommand{\eeq}[1]{\label{#1}\end{equation}}
\newcommand{\bea}{\begin{eqnarray}}
\newcommand{\eea}[1]{\label{#1}\end{eqnarray}}
\renewcommand{\Re}{\mbox{Re}\,}
\renewcommand{\Im}{\mbox{Im}\,}
\begin{titlepage}
\hfill  hep-th/0207135
\begin{center}
\hfill
\vskip .4in
{\large\bf N=1 NO-SCALE SUPERGRAVITY FROM IIB ORIENTIFOLDS}
\end{center}
\vskip .4in
\begin{center}
{\large S. Ferrara$^a$ and M. Porrati$^{b,c}$}\footnotemark
\footnotetext{e-mail: sergio.ferrara@cern.ch,
massimo.porrati@nyu.edu}
\vskip .1in
(a){\em Theory Division CERN, Ch 1211 Geneva 23, Switzerland}
\vskip .1in
(b){\em Department of Physics, NYU, 4 Washington Pl.,
New York, NY 10003, USA}
\vskip .1in
(c){\em Rockefeller University, New York, NY
10021-6399, USA}
\vskip .1in
\end{center}
\vskip .4in
\begin{center} {\bf ABSTRACT} \end{center}
\begin{quotation}
\noindent
We describe the low-energy effective theory of N=1 spontaneously broken 
supegravity obtained by flux-induced breaking in the presence of $n$ D3 branes.
This theory can be obtained by integrating out three massive gravitino 
multiplets in the hierarchical breaking 
${\rm N=4}\rightarrow 3 \rightarrow 2\rightarrow 1$ of a N=4 orientifold. 
This integration also eliminates the IIB complex dilaton.
The resulting theory is a no-scale supergravity model, whose moduli are the 
three chiral multiplets that correspond to the three radii of 
$T_2\times T_2\times T_2$ in $T_6$, together with the $6n$ brane coordinates.
The $U(n)$ gauge interactions on the branes respect the no-scale structure,
and the N=1 goldstino is the fermionic partner of the $T_6$ volume.
\end{quotation}
\vfill
\end{titlepage}
\eject
\noindent
\section{Introduction}
Type IIB orientifolds
with flux-induced supersymmetry breaking~\cite{fp,kst} offer 
interesting examples of models with moduli stabilization~\cite{ps,tv,m,m2} 
and hierarchical supersymmetry breaking with vanishing cosmological constant 
at the classical level.

They can be regarded as N-extended no-scale supergravities~\cite{cfkn,elnt} 
as far as the 
supersymmetry breaking deos not stabilize all moduli, thus offering the
possibility of creating a large hierarchy of scales~\cite{gkp}.

String and M--theoretical constructions of these models have been 
established~\cite{cv,bb,bw} but 
their low-energy description is only known in particular cases. Most of them 
involve ${\rm N=2 \rightarrow N=1}$ partial supersymmetry breaking, as in the 
case of models with R-R fluxes in Calabi-Yau 
compactifications~\cite{tv,m2,cklt,lm,ckkl,hl,da}. 

Recently, the effective, spontaneously broken  N=4 supergravities 
corresponding to a type IIB orientifold with arbitrary fluxes have been 
investigated~\cite{adfl2,dfv}. 
They have $0\leq {\rm N}' < 4$ surviving supersymmetries.
Those Lagrangians offer the possibility to obtain effective theories with
lower supersymmetry, by integrating out massive multiplets. For example,
the N=3 effective theory, obtained by integrating out a single massive 
gravitino multiplet is completely predicted by supersymmetry, and its symmetry
breaking terms arise by the gauging of its (Abelian) isometries~\cite{tz}. 
These 
isometries are related to the R-R scalars coming from the type IIB four 
form~\cite{fp,kst}.

In this paper, we would like to exhibit the particular simple form of the 
${\rm N=1\rightarrow  N=0}$ supergravity action, inclusive of D3 brane 
non-Abelian gauge interactions, and show that it provides, as expected, an 
exact no-scale N=2 supergravity model of the kind constructed long ago in the 
literature~\cite{cfkn,elnt,bcf}. 

The main distinction from the case previously studied, is that the 
supersymmetry breaking does not involve the IIB complex dilaton, $S$, since 
that field can be integrated out already at the level of the 
${\rm N=4\rightarrow N=3}$ breaking.
Therefore, the effective theory contains just the three chiral multiplets 
associated with the three radii (volumes) of $T_2\times T_2\times T_2=T_6$,
together with the brane coordinates, the $U(n)$ gauge couplings, and the flux
parameter $\mu$.
\section{Integrating out Massive Multiplets}
The apparently puzzling feature arising from integrating out the massive 
gravitino multiplets, in the presence of $n$ D3 branes, is that the scalar
manifold of the truncated theory is {\em not} a submanifold of the original 
theory, as it is, instead, in the absence of D3 branes, when the original 
theory contains only moduli coming from the bulk, 10-d sector~\cite{adf,l}.
The reason for this phenomenon is that the massless modes of the brane couple
to massive bulk fields. So, for instance, if $A^m_\mu$ is a massive bulk field,
the truncated theory in the presence of branes is not obtained by setting
$A^m_\mu=0$, but rather by setting $A^m_\mu\sim \phi \partial_\mu \phi$, where
$\phi$ denotes some brane coordinate. The effect of this procedure is a 
distortion of the original manifold, rather than a submanifold~\footnote{One 
could say~\cite{l} that the manifold in Eq.~(\ref{5}) is obtained by modding 
out an $R^6$ isometry of the manifold in Eq.~(\ref{4}), where the $R^6$ is 
gauged by the six vectors that become massive.}.
This phenomenon was implied by the discussion of Frey and 
Polchinski~\cite{fp}, in the
derivation of the ${\rm N=3}$ sigma-model metric in the presence of D3 brane
coordinates. It becomes 
evident by writing out the gauge covariant derivative of the R-R axions,
\beq
b^{IJ}=(b^{ij},b^{i \bar{\jmath}}), \qquad I=(i,\bar{\imath}), \; 
J=(j,\bar{\jmath}).
\eeq{1}
 The covariant derivative reads
\beq
D_\mu b^{ij}=\partial_\mu b^{ij} + C^{[i}_I \partial_\mu C^{j]}_I +
\epsilon^{ijk}A_k,
\eeq{2}
and results in the elimination of the term 
\beq
 \left|C^{[i}_I \partial_\mu C^{j]}_I\right|^2.
\eeq{3}
Also, the D3 brane action term
\beq
g_{ij}\partial_\mu 
C^i_I \partial^\mu C^j_I, \qquad g_{IJ}=(g_{ij},g_{i\bar{\jmath}}),
\eeq{3a}
is eliminated altogether from the two-derivative action, when, in the 
presence of fluxes, the $g_{ij}$ components of the metric acquire a mass term
(from the potential).
The effect of these changes is that the original N=4 sigma model 
\beq
{SU(1,1)\over U(1)} \times {SO(6,6+n)\over SO(6)\times SO(6+n)},
\eeq{4}
becomes~\cite{c&al}, 
after integrating out the scalar partners of the massive gravitino~\footnote{
The complex type IIB dilaton is part of the massive N=3 gravitino multiplet. 
This is the reason why the $SU(1,1)/U(1)$ factor drops out after integrating 
out the massive modes.},
\beq
{U(3,3+n)\over U(3)\times U(3+n)}.
\eeq{5}
This is not a submanifold of the former, unless $n=0$. 

If one further integrates out the second gravitino multiplet, in 
the presence of D3 branes, one finds, using the same reasoning,
\beq
{U(1,1+n)\over U(1)\times U(1+n)} \times {U(2,2+n)\over U(2)\times U(2+n)},
\qquad \mbox{N=2}
\eeq{6}
Here, the first factor is the manifold of vector multiplets, and the second
is the manifold of hypermultiplets.
By integrating out also the third gravitino multiplet (and a chiral multiplet)
one arrives at a N=1 theory with scalar manifold
\beq
{U(1,1+n)\over U(1)\times U(1+n)} \times {U(1,1+n)\over U(1)\times U(1+n)}
\times {U(1,1+n)\over U(1)\times U(1+n)} ,
\qquad \mbox{N=1}.
\eeq{6b}

Let us confine ourselves to the N=1 theory, with a residual gravitino mass 
$\mu$, related to the flux that breaks N=1 to N=0. The K\"ahler potential
for the manifold in Eq.~(\ref{6b}) is
\beq
K=\sum_{i=1}^3 K_i=-\sum_{i=1}^3 \log(t_i+\bar{t}_i -C_I^i \bar{C}_I^i).
\eeq{7}
If we turn on the $U(n)$ gauge interactions for the $n$ D3 branes, the 
superpotential becomes
\beq
W=\mu + f^{IJK} C_I^i C_J^j C_K^k \epsilon_{ijk},\qquad I=1,..,n^2,
\eeq{8}
while the D term is
\beq
D^K = \sum_{i=1}^3 {1\over t_i+\bar{t}_i -C_I^i \bar{C}_I^i} C_I^if^{IJK} 
\bar{C}^i_J.
\eeq{9}
In the two previous equations, $f^{IJK}$ are the $SU(n)$ structure constants, 
and $\mu$ is the flux~\footnote{In supergravity, this is a free parameter, but
string theory imposes on it a quantization condition.}
This theory is a no-scale model with scalar potential
\bea
V&=& \sum_{i=1}^3 e^K \left[ K^{C^i_I \bar{C}^i_J} {\partial W\over\partial 
C^i_I} {\partial \bar{W}\over \partial\bar{C}^i_J} \right] + 
\sum_{I} (D^I)^2 \nonumber \\
&=& e^K \left[ \sum_{i=1}^3 {1\over t_i+\bar{t}_i -C_I^i \bar{C}_I^i} 
{\partial W\over \partial C^i_I} {\partial \bar{W}\over \partial\bar{C}^i_J} 
\right] + \sum_{I} (D^I)^2.
\eea{10}
The gravitino mass term is
\beq
e^{K/2} |W| = \prod_{i=1}^3 (t_i+\bar{t}_i -C_I^i \bar{C}_I^i)^{-1/2}
| \mu + f^{IJK} C_I^i C_J^j C_K^k \epsilon_{ijk}|.
\eeq{11}
The vacuum is at $V=0$, i.e. $\partial W/\partial C^i_I=0$. Therefore, all $C$s
belong to the Cartan algebra of $U(n)$, and the gravitino mass is
\beq
m=\mu {1\over R_1 R_2 R_3},
\eeq{12}
with $R_i$ the radius of the $i$-th 2-torus in $T_2\times T_2 \times T_2$.
\section{Computation of the Scalar Potential}
In this 
Section we collect all formulas used to derive Eqs.~(\ref{10},\ref{11}).
\bea
V &=& e^K\left[ K^{a\bar{a}}D_aW \bar{D}_{\bar{a}}\bar{W} -3|W|^2\right],
\nonumber \\
K^{a\bar{a}}&=& (K)^{-1}_{a\bar{a}}; \qquad  K_{a\bar{a}}=\partial_a 
\partial_{\bar{a}}K.
\eea{13}
$K$ is given in Eq.~(\ref{7}), $W$ is given in Eq.~(\ref{8}).

For each $K_i$ we have

\bea
K_{t_i \bar{t}_i}&=& (t_i+\bar{t}_i -C_I^i \bar{C}_I^i)^{-2}, \qquad
K_{t_i}= -(t_i+\bar{t}_i -C_I^i \bar{C}_I^i)^{-1}, \nonumber \\
K_{t\bar{C}^i_I} &=& -C^i_I (t_i+\bar{t}_i -C_I^i \bar{C}_I^i)^{-2}, \qquad
K_{\bar{C}^i_I}=C^i_I (t_i+\bar{t}_i -C_I^i \bar{C}_I^i)^{-1}, \nonumber \\
K_{C^i_I\bar{C}^i_J}&=&{\delta_{IJ} \over 
t_i+\bar{t}_i -C_I^i \bar{C}_I^i} + {\bar{C}^i_I C^i_J \over 
(t_i+\bar{t}_i -C_I^i \bar{C}_I^i)^2}.
\eea{14} 
The entries of the inverse K\"ahler metric are
\bea
K^{t_i \bar{t}_i}&=& (t_i+\bar{t}_i -C_I^i \bar{C}_I^i)(t_i+\bar{t}_i),
\nonumber \\
K^{t\bar{C}^i_I}&=& \bar{C}^I_i (t_i+\bar{t}_i -C_I^i \bar{C}_I^i), \nonumber\\
K^{C^i_I\bar{C}^i_J}&=& \delta^{IJ} (t_i+\bar{t}_i -C_I^i \bar{C}_I^i).
\eea{15}
Notice the important identity 
\beq         
K_{t_i}K^{t\bar{C}^i_I}= - K_{C^i_J}K^{C^i_J\bar{C}^i_I}=-\bar{C}_I^i.
\eeq{15b}
In computing the scalar potential, one must remark that the terms proportional
to $|W|^2$ in $D_aW\bar{D}_{\bar{a}}\bar{W}K^{a\bar{a}}$ cancel the gravitino
mass term $-3|W|^2$. The cross terms proportional to 
$\bar{W} \partial W /\partial C^i_I$, cancel out because of Eq.~(\ref{15b}).
\section{Symmetries of the Model}
The no-scale supergravity so far described has some global symmetries besides
the obvious $U(n)$ gauge symmetry.

For $\mu=0$, even in the presence of gauge interactions, 
the theory is invariant under T-duality transformations
\beq
t_i\rightarrow 1/t_i, \qquad C^i_I \rightarrow {1\over t_i} C^i_I.
\eeq{16}
This symmetry is lost for $\mu\neq 0$, unless $\mu$ is replaced with
$\mu/ t_1 t_2 t_3$.

For any $\mu$, the theory is always invariant under 
$t_i \rightarrow t_i + i\beta_i$, and $C^i_I \rightarrow \exp(i\alpha^i)C^i_I$,
$\alpha_i,\beta_i\in R$, $\sum_{i=1}^3 \alpha^i=0$.

Notice that the above symmetries are sensitive to the cubic form of the the 
superpotential.  On the other hand, the no-scale structure of the model only 
depends on the particular form of the K\"ahler potential, as long as 
$W$ is independent of the moduli $t_i$. Some higher order $\alpha'$ corrections
have recently been computed, and they seem to affect the no-scale structure of
these models~\cite{bbhl}. 

We finally notice that the ${\rm N=1 \rightarrow N=0}$ phase is the only one 
that  does not result in a loss of moduli. Indeed, starting with 21 metric
deformations in the N=4 theory (without fluxes), 12 are lost in the
${\rm N=4\rightarrow N=3}$ breaking, together with the complex dilaton.
Four more metric 
moduli are lost in the breaking ${\rm N=3\rightarrow N=2}$, and two
more metric moduli are lost in the ${\rm N=2\rightarrow N=1}$ breaking. 
Therefore, in all
these phases, some of the metric moduli acquire a nontrivial potential. They 
precisely correspond to the superpartners of the charged axions. The latter,
are absorbed by the vector particles by the Higgs mechanism~\cite{adfl}.
The $n$ D3 brane coordinates, together with their superpartners, are
moduli in all phases, independently of the value of the four gravitino masses.
\section{Conclusions}
The interesting feature of the supergravity effective action of bulk fields
coupled to $n$ branes, is that it allows us to compute terms due to the 
gravitational back-reaction of the probe on the metric, as well as terms due 
to the non-abelian nature of the probe-brane action~\cite{m3}.
These effects, which are taken into account automatically by the requirement 
of local supersymmetry, may be very hard to compute in a string 
calculation. 

Interestingly enough, it turns out that these effects, at least in the 
approximation we are working on (two-derivative action), respect the exact
no-scale structure, including the non-abelian interactions of the D3 branes.

An example of these effects is that the $U(1)$ part of the $U(n)$ gauge group,
which decouples in the flat limit, has non-trivial gravitational couplings to
the other modes, because it enters in the K\''ahler potential. The 
contribution of the non-Abelian brane coordinates to the gravitino mass term,
\beq
e^{K/2}\psi_\mu \gamma^{\mu\nu}\psi_\nu 
f^{IJK}C^i_I C^j_J C^k_K \epsilon_{ijk},
\eeq{17}
is also another effect of the bulk-brane couplings requested by local
supersymmetry. 

Needless to say, these theories may also possess 
vacua where $\partial_C W\neq 0$. If this is the case, these vacua have a 
positive cosmological constant so that they give rise to de Sitter spaces.   
\vskip .1in 
\noindent
{\bf Acknowledgements}\vskip .1in
\noindent
M.P. would like to thank CERN for its kind hospitality. M.P. 
is supported in part by NSF grant no. PHY-0070787.
S.F. is supported in part by the European Community's Human Potential Program 
under contract HPRN-CT-2000-00131 Quantum Space-Time, and by DOE under grant
DE-FG03-91ER40662, Task C.

\end{document}